# A multiple-frame approach of crop yield estimation from satellite remotely sensed data

SUMANTA K. DAS*†, RANDHIR SINGH†
† Indian Agricultural Statistics Research Institute, Library Avenue,
New Delhi—110012

Many studies have recently explored the information from the satellite-remotely sensed data (SRSD) for estimating the crop production statistics. The value of this information depends on the aerial and spatial resolutions of SRSD. The SRSD with fine spatial resolution is costly and the aerial coverage is less. Use of multiple frames of SRSD in the estimation process of crop production can increase the precision. We propose an estimator for the average yield of wheat for the state of Haryana, India. This estimator uses the information from the Wide Field Sensor (WiFS) and the Linear Imaging Self Scanner (LISS-III) data from the Indian Remote Sensing satellite (IRS-1D) and the crop cutting experiment data collected by probability sampling design from a list frame of villages. We find that the relative efficiencies of the multiple-frame estimators are high in comparison to the single frame estimators.

## 1. Introduction:

India has a well established system for collecting agricultural statistics. The government′s economic policies depend on the production, area and yield statistics of different crops. The crop production statistic is the product of crop′s yield and its growing areas. Generally the area statistics is estimated by complete enumeration. Very few studies have used the remote sensing technology for crop area estimation (Dadhwal *et al.* (1987), Parihar *et al.* (1987), Das and Singh (2009)). The General Crop Estimation Surveys (GCES) program collects the crop cutting experiment (CCE) data through the multi-stage simple random sampling from a list frame of villages to estimate the average yield. There is immense scope to improve the reliability and timeliness of these estimates using the satellite-remotely sensed data (SRSD).

Even though the multiple-frame surveys (MFS) are most commonly used for the agricultural domains (Gonzalez-Villalobos and Wallace, 1996, National Agricultural Statistics Service, 1999), still many practitioners and researchers in this field have proposed several forms of MFS for social, establishment, business, household or individual-based surveys (Skinner (1991), Skinner *et al.*(1996)). A MFS is an approach to combine two or more sampling frames when one frame is complete but expensive to sample; other frames are inexpensive to sample but incomplete. However, in many cases, a frame that covers the entire population is very expensive to sample from. An alternate frame may be available that does not cover the entire population but cheaper to sample from. For example in an agricultural survey on wheat for the state of Haryana in India, an area frame (e.g. satellite image) may include all the wheat growing areas but selecting samples from this frame will increase the cost and complexity, because there is already one existing system for collecting the samples based on the list frame. A better option will be to combine the random samples taken from the list frame with the SRSD (as area frame) for estimation of the population quantities with higher precision.

---

* Now at the Institute For System Studies And Analyses, Defence Research And Development Organization, NewDelhi-110054, Email: sumantadas.delhi@ieee.org

In this study, we propose an estimator for average yield of a finite population considering the multiple-frame surveys for multi-stage sampling. The most common example of multiple-frame agricultural survey (Gonzalez-Villalobos and Wallace, 1996) combines the area sample component with the list sample component considering that samples belong to either any one of the frames, not in more than one frames. This aspect simplifies the estimation process. But this is not always the reality, in this study we consider the complex situation when the samples belong to more than one or to all frames. This estimation technique is based on the idea by Bankier (1996), Skinner *et al.* (1994) to view a multiple-frame sample as a special case of selecting two or more samples independently from the same frame.

The objective of this study is to apply the multiple-frame approach to improve the exiting list frame approach of the crop yield estimation. For this purpose, we combine the existing CCE data (collected by the GCES program) with the satellite image frames over the study area. We assume that the CCE data are also the random samples taken from the satellite image frames.

We have performed this study to estimate the average yield of wheat for the state of Haryana, India for the year 1997-98 using the Wide Field Sensor (WiFS) and Linear Imaging Self Scanner (LISS-III) data from the Indian Remote Sensing (IRS-1D) satellite. The basic aim is to measure the performance of the multiple-frame surveys for crop yield estimation through SRSD and CCE data. The purpose is to identify the advantages and complexity of the MFS over the traditional method before proposing for adapting it at the national level. In the recent times several countries are already conducting experiments to adopt the MFS approach for collecting their agricultural statistics.

## 2. Background

The most frequent applications of multiple-frame in agriculture combines the area frames with the list frames (Gonzalez-Villalobos and Wallace, 1996). As Lohr and Rao (2006) discussed area frame in comparison to list frame. Area frame is just the division of a map or an aerial photograph or a satellite image into a number of segments of land where segments are selected and studied by probability sampling. On the other hand list frames are generally names and addresses of agricultural farmers.The area frame is complete and insensitive to changes, but very expensive to sample. The list frames are usually less costly to sample, but the lists may not include all units. In such studies frames are assumed to be non-overlapping. A multiple-frame estimate is just the summation of the estimates from these two frames. Complexity arises when the sampling frames are overlapped. Although various theoretical studies have been performed for improving the multiple-frame estimates in different circumstances, but very few studies have been performed to apply the actual overlapped frames for agricultural surveys using SRSD.

Hartlay in 1962 derived the basic theory of MFS for two overlapping frames. Two basic assumptions of his theory were the completeness (i.e. every unit in the population must belong to at least one of the frames) and the identifiability (i.e. the belongingness of a sampled unit to a frame is known). Following the general convention of defining the multiple-frame set up, figure 1 shows two different instances of MFS. In figure 1(*a*), three frames $A_1$, $A_2$, and $A_3$ each frame is incomplete and together with some overlap they form the entire population $\Omega$ of size *N*. These two frames form $2^3-1$ mutually exclusive domains ({*1*},{*2*},{*3*},{*1, 2*},{*1, 3*}, {*2, 3*} and {*1, 2, 3*}). The figure 1(*b*) is also a three-frame design but has four domains, because frame $A_1$ is complete and frames $A_2$ and $A_3$ are incomplete but overlapping. Because the frame $A_1$ is complete.

Since the pioneering work of Hartley (1962) on MFS, we have seen various estimators have been proposed for the estimating the population parameters (like mean, total and variances). The estimators of Hartley (1962, 1974) and Fuller-Burmeister (1972) are the best linear unbiased estimators among the class of other linear unbiased estimators of the population total (*Y*) because they have minimum variances. The drawbacks of these estimators are the different set of weights ($w_i$) associated with each response variable $y_i$. To obtain these weights, we need to minimize the variance expressions that lead the estimators non-linear functions of $y_i$'s. Bankier (1986) considered that the estimation of MFS could be viewed as special case of selecting two or more samples independently from the same frame. He showed that "standard technique from literatures for estimating from a single frame, such as Horvitz-Thompson estimator or ratio estimation could be applied to MFS." Bankier (1986), Kalton-Anderson (1986) developed the estimators of *Y* that uses only single formula of $w_i$'s . Generally speaking these weights are proportional to the inclusion probabilities of the sampling units.

Hartley (1962) considered only the situation for simple random sampling. Afterwards the estimators of Fuller-Burmeister (1972), Hartley (1974), Bankier (1986) and Skinner (1991) have considered the general case when at least one of the samples is selected by a complex design, e.g., multi-stage random sampling. Saxena *et al*. (1986) extended the Hartley's estimator for two-stage sampling. They have discussed how the complexity of the MFS for multi-stage increases as the alternatives for multiple-frame approach multiply, e.g. in two-stage sampling there may be multiple frames at the first stage and single frame at the second stage or vice-versa. They also derived the optimal variance expression for multiple frames at both stages and studied the gain in efficiency due to application of multiple frames instead of single frame at both the stages. The problem of different set of weights still remains with these estimators.

Bankier (1986) first applied the raking ratio estimator for dual frame surveys and found it more efficient than the Hartley′s estimator but he has not given any theoretical justification. Skinner (1991) theoretically studied and justified the efficiency of the raking ratio estimator. Skinner *et al.* (1994) have used single frame estimation in the context of multivariate stratification. He proposed to select independent subsamples by stratified random sampling with respect to each stratification variable related to the corresponding study variables. He then combined the different subsamples using the estimation technique of MFS.

Rao (1983) showed that the unbiased estimate of the population total can be derived from the maximum likelihood principles when a simple random sample is taken from each frame. Skinner and Rao (1996) discussed the internal inconsistency of the estimators proposed by Hartley (1962), Fuller and Burmeister (1972) due to different set of weights for each response variable $y_i$. Skinner and Rao (1996) proposed the pseudo-maximum likelihood estimator (PMLE) for complex design like multi-stage for dual frame surveys. They studied the asymptotic properties of the PMLE and found it similarly efficient as Hartley's and Fuller-Burmeister's estimators. They also proposed the single frame estimator for complex design. They argued that the single frame estimator has advantages due to same set of weights to each response variable. Lohr and Rao (2000) compared the asymptotic efficiencies of these estimators in dual frame surveys and found that the PMLE combined high efficiency with applicability to complex surveys. Variance estimation is cumbersome for MFS. Lohr and Rao (2000) discussed the Taylor linearization and jackknife methods. Lohr and Rao (2007) proposed the bootstrap method for variance estimation.

Very recently Lohr and Rao (2006) have proposed the PMLE for complex MFS introducing the general matrix notations for expressing various estimators and extended the Skinner and Rao's (1996) dual frame PMLE for multiple-frame. After a simulation study they have found that the PMLE performs better than the single frame estimator as well as the raking ratio estimator provided sample size is very large, and so, they have concluded that the PMLE is the best estimator because it has internal consistency as well as asymptotic variance property under the condition of very large sample size. They have recommended to use either the PMLE or the single frame estimator with the raking ratio adjustment when the number of frames, $\alpha$ is greater than 2, also they have added when $\alpha \geq 3$, the theoretically optimal Fuller-Burmeister and Hartley's methods become unstable, because it requires solving system of equations using a large estimated covariance matrix.

Very few studies have been performed to apply various sampling techniques on SRSD for agricultural crop surveys. Singh *et al.* (1992, 2002), Singh (2003) have illustrated that the stratification based on the spectral vegetation index improved the efficiency of crop yield estimation significantly. In those studies post-stratified estimators have been used to combine the CCE data and the stratified satellite data for crop yield estimation.

## 3. Sampling Design

Let the finite population of interest, $\Omega$ is the union of $\alpha$ frames. The frames may overlap and be complete or incomplete. If all the frames are incomplete resulting in a possible $2^{\alpha} - 1$ nonoverlapping domains. When $\alpha = 2$, it is called a dual-frame surveys. Let we assume that each frame consists of $N_{\alpha}$ ($\alpha = 1...A$) primary stage units (psu's), the $i_{\alpha}^{th}$ psu comprising of $M_{\alpha_i}$ second stage units (ssu's). The $n_{\alpha}$ psu's are selected with simple random sampling without replacement (srswor) at the first stage and from the $i_{\alpha}^{th}$ selected psu, $m_{\alpha_i}$ ssu's are selected with srswor. Let we define,

$Y_{\alpha ij}$ ($y_{\alpha ij}$) = the value of characteristic, $y$ on the $j_{\alpha}^{th}$ ssu in the $i_{\alpha}^{th}$ (selected) psu with respect to the frame $\alpha$ in the population (sample). $j_{\alpha} = 1,…, M_{\alpha_i}$ ($m_{\alpha_i}$); $i_{\alpha} = 1,…, N_{\alpha}$ ($n_{\alpha}$). In the present study, $Y_{\alpha ij}$ represent the average yield of the village (the CCE data). It also represents the average yield of pixel on the satellite image frame.

Our population parameter of interest is the average yield of the state of Haryana ($\bar{\bar{Y}}$). We propose an unbiased estimator of the average yield of the population based on the sample observations as (see next section for proof):

$$\hat{\bar{\bar{y}}} = \sum_{\alpha=1}^{A} \sum_{i_\alpha=1}^{n_\alpha} \sum_{j_\alpha=1}^{m_{\alpha_i}} w_{\alpha ij} y_{\alpha ij} \tag{1}$$

where
$$w_{\alpha ij} \propto \left( \sum_{\alpha=1}^{A} \pi_\alpha(i,j) \right)^{-1} \tag{2}$$

$\pi_\alpha(i,j)$ is the inclusion probability (the probability of including the $y_{\alpha ij}{}^{th}$ observation in the sample). We define this as:

$$\pi_\alpha(i,j) = P(i \in n_\alpha \text{ and } j \in m_{\alpha_i}) = \left( \frac{n_\alpha}{N_\alpha} \frac{m_{\alpha_i}}{M_{\alpha_i}} \right). \tag{3}$$

The estimator has the following variance expression (see next section for proof)

$$V(\hat{\bar{\bar{Y}}}) = \sum_{\alpha=1}^{A} \left\{ \overline{m}_\alpha (1 - f_{\alpha_1}) . S'^2_{\alpha b} + \frac{n_\alpha}{N_\alpha} \sum_{i_\alpha=1}^{N_\alpha} (1 - f^2_{\alpha 2_i}) S^2_{\alpha w_i} \right\} \tag{4}$$

where $f_{\alpha_1} = \dfrac{n_\alpha}{N_\alpha}$, $f_{\alpha 2_i} = \dfrac{m_{\alpha_i}}{M_{\alpha_i}}$, $\overline{m}_\alpha = \dfrac{1}{n_\alpha} \sum_{i_\alpha=1}^{n_\alpha} m_{\alpha_i}$ and

$$S'^2_{\alpha b} = \frac{1}{(N_\alpha - 1)} \sum_{i_\alpha=1}^{N_\alpha} (\overline{Z}_{\alpha i} - \overline{\overline{Z}}_{\alpha..})^2 \tag{5}$$

$$S^2_{\alpha w_i} = \frac{1}{(M_{\alpha_i} - 1)} \sum_{j_\alpha=1}^{M_{\alpha_i}} (Z_{\alpha ij} - \overline{Z}_{\alpha i})^2 \tag{6}$$

$$Z_{\alpha ij} = w_{\alpha ij} . Y_{\alpha ij}, \quad \overline{Z}_{\alpha i.} = \frac{\sum_{j_\alpha=1}^{M_{\alpha_i}} w_{\alpha ij} . Y_{\alpha ij}}{M_{\alpha_i}} \text{ and } \overline{\overline{Z}}_\alpha = \frac{\sum_{i_\alpha=1}^{N_\alpha} M_{\alpha_i} . \overline{Z}_{\alpha i}}{N_\alpha}.$$

We define an unbiased estimator (see next section for proof) of variance based on the sample observations as:

$$Est(V(\hat{\bar{\bar{y}}})) = \sum_{\alpha=1}^{A} \left\{ \overline{m}_\alpha (1 - f_{\alpha_1}) s'^2_{\alpha b} + \frac{n_\alpha}{N_\alpha} \sum_{i_\alpha=1}^{n_\alpha} (1 - f^2_{\alpha 2_i}) s^2_{\alpha w_i} \right\} \tag{7}$$

where

$$s'^2_{\alpha b} = \frac{1}{(n_\alpha - 1)} \sum_{i_\alpha=1}^{n_\alpha} (\overline{z}_{\alpha i.} - \overline{\overline{z}}_{\alpha..})^2 \tag{8}$$

$$s^2_{\alpha w_i} = \frac{1}{(m_{\alpha_i} - 1)} \sum_{j_\alpha=1}^{m_{\alpha_i}} (z_{\alpha ij} - \overline{z}_{\alpha i})^2 \tag{9}$$

$$z_{\alpha ij} = w_{\alpha ij} . y_{\alpha ij}, \quad \overline{z}_{\alpha i.} = \frac{\sum_{j_\alpha=1}^{m_{\alpha_i}} w_{\alpha ij} . y_{\alpha ij}}{m_{\alpha_i}} \text{ and } \overline{\overline{z}}_{\alpha..} = \frac{\sum_{i_\alpha=1}^{n_\alpha} m_{\alpha i} . \overline{Z}_{\alpha i}}{n_\alpha}$$

We present a list of standard notations used for MFS in table 1.

## 4. Proof of unbiasedness and varience expression

Let we define $E_1$, $V_1$ respectively, the expectation and variance operators of a statistics with respect to primary stage units (psu's) and $E_2, V_2$ denote respectively expectation and variance with respect to second–stage samples for a given sample of psu's.

$$E(\bar{\bar{y}}) = E_1 E_2 (\sum_{\alpha=1}^{A} \sum_{i_\alpha=1}^{n_\alpha} \sum_{j_\alpha=1}^{m_{\alpha_i}} w_{\alpha ij} y_{\alpha ij})$$

$$= E_1 (\sum_{\alpha=1}^{A} \sum_{i_\alpha=1}^{n_\alpha} \sum_{j_\alpha=1}^{m_{\alpha_i}} w_{\alpha ij} E_2(y_{\alpha ij} / j_\alpha)$$

$$= E_1 (\sum_{\alpha=1}^{A} \sum_{i_\alpha=1}^{n_\alpha} \sum_{j_\alpha=1}^{m_{\alpha_i}} w_{\alpha ij} \frac{\sum_{j_\alpha=1}^{M_{\alpha_i}} Y_{\alpha ij}}{M_{\alpha_i}})$$

$$= E_1 (\sum_{\alpha=1}^{A} \sum_{i_\alpha=1}^{n_\alpha} \sum_{j_\alpha=1}^{m_{\alpha_i}} w_{\alpha ij} \bar{Y}_{\alpha_i})$$

$$= \sum_{\alpha=1}^{A} \sum_{i_\alpha=1}^{n_\alpha} \sum_{j_\alpha=1}^{m_{\alpha_i}} w_{\alpha ij} E_1(\bar{Y}_{\alpha_i.} / i_\alpha)$$

$$= \sum_{\alpha=1}^{A} \sum_{i_\alpha=1}^{n_\alpha} \sum_{j_\alpha=1}^{m_{\alpha_i}} w_{\alpha ij} \frac{\sum_{i_\alpha=1}^{N_\alpha} \bar{Y}_{\alpha_i}}{N_\alpha}$$

$$= \bar{\bar{Y}}, \text{ if and only if } \sum_{\alpha=1}^{A} \sum_{i_\alpha=1}^{n_\alpha} \sum_{j_\alpha=1}^{m_{\alpha_i}} w_{\alpha ij} = 1,$$

Hence, the sample mean ($\bar{\bar{y}}$) is an unbiased estimator of the population mean ($\bar{\bar{Y}}$).

Again, Let we assume,

$$Z_{\alpha ij} = w_{\alpha ij} Y_{\alpha ij}, \quad \bar{Z}_{\alpha_{i.}} = \frac{\sum_{j_\alpha=1}^{M_{\alpha_i}} Z_{\alpha ij}}{M_{\alpha_i}} \quad \text{and} \quad \bar{\bar{Z}}_{\alpha_{..}} = \frac{\sum_{i_\alpha=1}^{N_\alpha} \bar{Z}_{\alpha_{i.}} M_{\alpha_i}}{N_\alpha},$$

We know for the two-stage random sampling (Cochran (1977), Mukhopadhyay (1998)),

$$V(\bar{\bar{y}}) = E_1 V_2(\bar{\bar{y}}) + V_1 E_2(\bar{\bar{y}})$$

Now

$$E_1 V_2 (\sum_{\alpha=1}^{A} \sum_{i_\alpha=1}^{n_\alpha} \sum_{j_\alpha=1}^{m_{\alpha_i}} Z_{\alpha ij} / j_\alpha)$$

$$= E_1 (\sum_{\alpha=1}^{A} \sum_{i_\alpha=1}^{n_\alpha} \sum_{j_\alpha=1}^{m_{\alpha_i}} \left( \frac{1}{m_{\alpha_i}} - \frac{1}{M_{\alpha_i}} \right) S^2_{w_{\alpha_i}})$$

$$= \sum_{\alpha=1}^{A} \frac{n_\alpha}{N_\alpha} \sum_{j_\alpha=1}^{N_\alpha} (1 - f_{\alpha 2_i}) S^2_{w_{\alpha_i}}$$

and

$$V_1 E_2 (\sum_{\alpha=1}^{A} \sum_{i_\alpha=1}^{n_\alpha} \sum_{j_\alpha=1}^{m_{\alpha_i}} Z_{\alpha ij})$$

$$= \sum_{\alpha=1}^{A} \sum_{i_\alpha=1}^{n_\alpha} \sum_{j_\alpha=1}^{m_{\alpha_i}} V_1(\bar{Z}_{\alpha_i})$$

$$= \sum_{\alpha=1}^{A} \sum_{i_\alpha=1}^{n_\alpha} \sum_{j_\alpha=1}^{m_{\alpha_i}} \left( \frac{1}{n_\alpha} - \frac{1}{N_\alpha} \right) S'^2_{\alpha b}$$

$$= \sum_{\alpha=1}^{A} \left( m_{\alpha_i} (1 - f_{\alpha_1}) S'^2_{\alpha b} \right)$$

$$\cong \sum_{\alpha=1}^{A} \left( \bar{m}_\alpha (1 - f_{\alpha_1}) S'^2_{\alpha b} \right),$$

$$\text{where } \bar{m}_\alpha = \frac{1}{n_\alpha} \sum_{i_\alpha=1}^{n_\alpha} m_{\alpha_i}$$

Hence

$$V(\bar{\bar{y}}) = \sum_{\alpha=1}^{A} \left\{ \bar{m}_\alpha (1 - f_{\alpha_1}) S'^2_{\alpha b} + \frac{n_\alpha}{N_\alpha} \sum_{j_\alpha=1}^{N_\alpha} (1 - f_{\alpha 2_i}) S^2_{w_{\alpha_i}} \right\}, \text{hence it proves the equation}$$

(4).

Now we derive the unbiased estimate of the variance of sample mean (equation 7)

$$E((n_\alpha - 1)s^2_{\alpha b}) = E(\sum_{i_\alpha=1}^{n_\alpha} (\bar{z}_{\alpha_i} - \bar{\bar{z}}_{\alpha..})^2)$$

$$= E_1 E_2 \left[ \sum_{i_\alpha=1}^{n_\alpha} \bar{z}^2_{\alpha_i} - \frac{\left( \sum_{i_\alpha=1}^{n_\alpha} \bar{z}_{\alpha_i} \right)^2}{n_\alpha} \right] \qquad (10)$$

Now,

$$E_1 E_2 \left[ \sum_{i_\alpha=1}^{n_\alpha} \bar{z}^2_{\alpha_i} \right]$$

$$E_1 \left[ \sum_{i_\alpha=1}^{n_\alpha} \left\{ \bar{z}^2_{\alpha..} + \frac{(1 - f_{\alpha 2_i}) S^2_{\alpha w_i}}{m_{\alpha_i}} \right\} \right]$$

$$= n_\alpha \sum_{i_\alpha=1}^{N_\alpha} \frac{\bar{Z}^2_{\alpha..}}{N_\alpha} + n_\alpha \sum_{i_\alpha=1}^{N_\alpha} \frac{1 - f_{\alpha 2_i}}{m_{\alpha_i}} \cdot \frac{S'^2_{\alpha w_i}}{N_\alpha} \qquad (11)$$

Again

$$E \left( \sum_{i_\alpha=1}^{n_\alpha} \bar{z}_{\alpha_i} \right)^2 = V \left( \sum_{i_\alpha=1}^{n_\alpha} \bar{z}_{\alpha_i} \right) + \left( E \left( \sum_{i_\alpha=1}^{n_\alpha} \bar{z}_{\alpha_i} \right) \right)^2$$

$$= n^2_\alpha V(\bar{\bar{z}}_{\alpha..}) + n^2_\alpha \cdot \bar{\bar{z}}^2_{\alpha..} \qquad (12)$$

Hence from the equations (11) and (12), the equation (10) reduces to

$$n_\alpha \sum_{i_\alpha=1}^{N_\alpha} \frac{\overline{Z}_{\alpha..}^2}{N_\alpha} + n_\alpha \sum_{i_\alpha=1}^{N_\alpha} \frac{1-f_{\alpha 2_i}}{m_{\alpha_i}} \cdot \frac{S_{\alpha w_i}^{'2}}{N_\alpha} + n_\alpha V(\overline{\overline{z}}_{\alpha..}) + n_\alpha \cdot \overline{\overline{z}}_{\alpha..}^2$$

$$Est(V(\overline{\overline{\overline{y}}})) = \sum_{\alpha=1}^{A} \left\{ \overline{m}_\alpha (1-f_{\alpha_1}) s_{\alpha b}^{'2} + \frac{n_\alpha}{N_\alpha} \sum_{j_\alpha=1}^{n_\alpha} (1-f_{\alpha 2_i}) s_{\alpha w_i}^2 \right\}$$

## 5. Study area and data preparations

The state of Haryana and the districts, Rhotak and Jhajjar, are one of the major wheat growing areas in India, having acreage of more than 50 percent under wheat crop during Rabi season (Oct. to Jan.). Geographically Haryana lies between 74°25′ to 77°38′ E longitude and 27°40′ to 30°55′ N latitude and the districts, Rhotak and Jhajjar, lie between 76°15′ to 77°00′ E longitude and 28°40′ to 29°05′ N latitude.

The image frames were acquired on 16$^{th}$ February 1998 for the path 30 and row 47 at 11 AM. The WiFS image covers the state of Haryana completely and the LISS III covers the districts, Rhotak and Jhajjar, in Haryana. The WiFS has a spatial resolution of 188 × 188 m$^2$ and two spectral bands, one in visible band i.e. 620-680 nm and another in infra red (IR) band i.e. 770-860 nm. Also its swath is 810 Km. Whereas LISS III has a spatial resolution of 23.5 × 23.5 m$^2$ and four spectral bands, two bands in visible region and two in IR region. The bands are 520-590 nm, 620-680 nm, 770-860 nm, and 1550-1700 nm and its swath is 148 kms. The roads, built up area, agricultural land are identifiable on these images.

The National Sample Survey Organization (NSSO), in India, collects the CCE data under the GCES from a list frame of villages. A multi-stage random sampling design is adopted in these surveys where the districts constitute the psu. The villages within the districts are ssu. A random sample of villages is selected from the selected districts. From each selected village, two fields are selected randomly and from each field, a plot of fixed size, generally measuring 10 × 5 m$^2$; is selected.

Figure 2 shows the boundaries of the state of Haryana along with its 20 districts. The districts, Rhotak and the Jhajjar, (shown with darker lines) were selected with srswor. These districts have been reported to consist of total 147 and 247 villages. We have drawn a lighter shaded line to represent the villages as ssu inside the selected psu′s. A second stage sample consists of 35 and 36 villages are selected from these two villages with srswor. The sampled villages are visited by the survey workers and two plots of size 10 × 5 m$^2$ are selected to calculate the average yield of the villages.

The geographical locations of each sampled fields are recorded by a differential Global Positioning System (GPS). A total of 25 ground control points (GCP) are selected from the well defined road intersections. The measurements are referenced to the European Datum 1950 (ED50) and the UTM-zone 36 projection. The GPS observations are made at the GCPs using an AshtechZ12 receiver. The horizontal measure of accuracy of this GPS is of 10 mm (Ashtech 1993). After completing GPS measurements, a handheld Magellan GPS 315 receiver is used in autonomous operation to measure the coordinates of the sampled fields. The Magellan GPS 315 is a single frequency Coarse/Acquisition (C/A) code receiver. It uses 12 parallel channels, working simultaneously, to locate and collect data from the GPS satellites. The rms error it provides is 15 m (Magellan Systems Corporation 1999). The coordinates derived from the autonomous GPS observations are automatically transformed to datum ED50 and UTM-zone 36 projection by the firmware of the receiver.

The geometric correction is carried out using the IDRISI image analysis software. The WiFS and LISS III images are geometrically corrected to the UTM zone 36 projection and to the ED50 datum using the 25 GPS derived GCPs. The geometric correction is based on the second-degree polynomial and the nearest neighbor resampling techniques (The RESAMPLE module of IDRISI is used). The locations collected by the GPS are used to identify the selected wheat fields on the imagery.

We have developed the boundary masks of the study areas using a topographic map of scale 1:50,000 and a digitizer. The digitized map is superimposed over the satellite imagery to extract all pixels belonging to the study areas. The False Color Composites (FCC′s) are generated using the bands : 2, 3, 4 (of LISS III) and 1, 2 (of WiFS) for the districts, Rhotak, Jhajjar, (Fig. 3(a)) and the state of Haryana (Fig. 3(b)) respectively.

## 6. Estimation

To establish the multi-stage sampling on the SRSD we need to form spectral clusters. In terms of remote sensing and pattern classification, clustering implies grouping of pixels without using any *a prior* information in the multispectral space. Clusters are generally made up of neighboring elements and pixels belonging to a particular cluster tend to have spectrally similar characteristics. As a simple rule, to make multi-stage sampling more efficient the elements in a cluster should be less and the number of clusters should be large.

In order to establish this relationship it is necessary to give the definition of clusters and their criterion. There are various criteria discussed by Fukunaga (1990), Duda *et al.* (2001) for clustering. We have used the *K*-means clustering method though it is enormously applied to SRSD and it has the flexibility to define the number of clusters. The *K*-means clustering is based on the criterion of minimizing the sum of square error defined as:

$$E = \sum_{i=1}^{N} \sum_{k=1}^{K} D^2(\mathbf{x}_i, \mathbf{z}_k) \tag{13}$$

where $N$ is the total number of pixels, $K$ is the total number of clusters, $x_i$ is the $i^{th}$ pixel measurement vector of *p*-dimensions, *p* is the number of bands in the SRSD and $\mathbf{z}_k$ is the vector of *K* cluster means and $D^2(x_i, z_k)$ is the Euclidean distance between the point $x_i$ and the cluster centers $z_k$.

This objective is achieved through an iterative optimization technique. Initial cluster means are randomly chosen to start the algorithm (usually selected from the domain space). Then iteratively the means are calculated so that the clustering criterion is minimized. The termination of the algorithm is decided by a factor ($\epsilon$, epsilon) which measures the successive differences of the cluster means. If the $\epsilon$ is lower than the pre-specified value then the algorithm terminates. Duda *et al.* (2001) stated that the estimates obtained through *K*-means clustering algorithm is the approximate maximum likelihood estimate of the clusters means. The order of complexity is *O(NpKT)* where *T* is the number of iterations (Duda *et al.* (2001)). We know from the sample survey theory that the within cluster sum of square should be maximum and between sum of square should be minimum to make the cluster sampling efficient (Cochran (1977), Mukhopadhyay (1998)). The criterion of *K*-means also produces the clusters that satisfy this theory.

From the list frame perspective, we can imagine that each CCE data is a realization of the average yield (a stochastic variable) of the village. Similarly, the village yield is a realization of the average yield of the districts. And district yield is a realization of the average yield of the state. On the otherhand, from the satellite frame perspective, each plot's yield is a realization of the average yield of the pixel. The pixel is a realization of the average yield of the spectrally similar clusters. Again the average yield of each cluster is a realization of the average yield of the entire population.

We have used the corrected raw bands satellite data for clustering. The satellite images are clustered using the *K*-means clustering algorithm in the MATLAB software. In both the frames different numbers of clusters are formed (500 for WiFS and 3000 for LISS III ). Figure 4(*a-b*) shows the clusters generated on the SRSD of the study areas. The rectangular box on the figure 4(*a*) represents the sampling zone of the CCE data. The scatter plot of the sampled fields is shown with their geographical locations (longitude and latitude) on figure 4(*c*).

The belongingness of each CCE data into these clusters were obtained by their GPS-collected locations. The belongingness (or allocation) of samples in different clusters are given in the table 2. We apply the proposed estimator (equation (1)) to both the single and the multiple-frame set up.

The weights are calculated using the equations (2) and (3). These are the sampling weights (inverses of the probabilities of selection), where each sampling weight is multiplied by 1/ (sum of sampling weights) so that the new weights sum to 1. We have assumed that we can correctly identify the domain membership of each sampling unit in each of the surveys. This is possible due to the availability of geographical locations of the sampled fields. This is an important assumption of this study. Let we define $\pi_\alpha(i)$ be the inclusion probability of the $i^{th}$ psu to be selected in the sample for the $\alpha^{th}$ frame and $\pi_\alpha(j|i)$ be the inclusion probability of the $j^{th}$ element in the $i^{th}$ psu. Then the overall inclusion probability of $j^{th}$ ssu to be in the $i^{th}$ psu is the product of $\pi_\alpha(i)$ and $\pi_\alpha(j|i)$, i.e. $\pi_\alpha(i,j) = \pi_\alpha(i) \times \pi_\alpha(j|i)$. Though at both the stages samples are selected by srswor, $\pi_\alpha(i) = \dfrac{n_\alpha}{N_\alpha}$ and $\pi_\alpha(j|i) = \dfrac{m_{\alpha_i}}{M_{\alpha_i}}$.

Hence $\pi_\alpha(i,j) = \pi_\alpha(i) \times \pi_\alpha(j|i) = \dfrac{n_\alpha}{N_\alpha} \times \dfrac{m_{\alpha_i}}{M_{\alpha_i}}$. The design weight is inversely proportional to the summation of the inclusion probabilities to different frames i.e.

$$w_{\alpha ij} \propto \left( \sum_{\alpha=1}^{A} \pi_\alpha(i,j) \right)^{-1} \text{ or } w_{\alpha ij} = \left( \sum_{\alpha=1}^{A} \pi_\alpha^*(i,j) \right)^{-1} \text{ where } \pi_\alpha^*(i,j) = \dfrac{1}{M_{\alpha_0}} \times \dfrac{n_\alpha}{N_\alpha} \times \dfrac{m_{\alpha_i}}{M_{\alpha_i}}$$

and $M_{\alpha_0}$ is the total number of ssu's in the $\alpha^{th}$ frame.

## 7. Results and Discussion

This study estimates the average yield of wheat for the state of Haryana, India combining three sampling frames (one list frame and two SRSD). Although we have taken only two SRSD for this experiment, a number of SRSD can be added in the estimation process. The suitability is dependent on their spatial resolutions. We have combined the existing crop cutting experiment (CCE) data with SRSD to estimate the average yield of wheat for the state of Haryana using the estimation technique of multiple-frame sampling (MFS). We have identified the belongingness of each CCE data (with the help of their geographical locations) into different stages of sampling units. We have used a single set of weights to each of the CCE data. We have used the *K*-means clustering algorithm for developing clusters on the SRSD. The clusters form the primary stage units. The cluster sizes are used for generating the weights.

Table 2 shows the estimation of the average yield using the conventional two-stage sampling estimator (i.e. $\hat{\bar{\bar{y}}}_\alpha = \dfrac{\sum_{i_\alpha=1}^{n_\alpha} M_{\alpha_i} \bar{y}_{\alpha_i}}{n_\alpha \bar{M}_\alpha}$ and

$$v(\hat{\bar{\bar{y}}}_\alpha) = \left(\dfrac{1}{n_\alpha} - \dfrac{1}{N_\alpha}\right) s'^2_{\alpha b} + \dfrac{1}{n_\alpha N_\alpha} \sum_{i_\alpha=1}^{n_\alpha} \dfrac{M_{\alpha_i}}{\bar{M}_\alpha}\left(\dfrac{1}{m_{\alpha_i}} - \dfrac{1}{M_{\alpha_i}}\right) s^2_{\alpha w_i},$$

where $\hat{\bar{\bar{y}}}_\alpha$ is the overall sample mean, $\bar{y}_{\alpha_i}$ is the sample mean of the $i_\alpha{}^{th}$ psu, $n_\alpha$ is the sample size of psu's, $N_\alpha$ is the population size of psu's, $m_{\alpha_i}$ is the sample size of the $i_\alpha{}^{th}$ ssu, $M_{\alpha_i}$ is the population size of the $i_\alpha{}^{th}$ ssu's, $\bar{M}_\alpha$ is the average population size of ssu's, $s'^2_{\alpha b}$ is the between mean sum of square of the sampled psu's and $s^2_{\alpha w_i}$ is the within mean sum of square of the sampled psu's and $\alpha$ is the number of sampling frames.

This table shows the estimated average yield of the state of Haryana from the conventional random samples of CCE data (total 71 sample points, same sample points are assumed to be the part of the image frames, thus $n_1 = n_2 = n_3 = 71$) from the list frame. We have estimated the average yield using the information from the list frame (3$^{rd}$ Column of the table 2) and the satellite frames (4$^{th}$ and 5$^{th}$ Columns of the table 2). The state of Haryna has total of 20 districts. We have divided the WiFS and LISS III data into 500, 3000 clusters (to keep the average size of the psu ($\bar{M}_\alpha$) to near 500) respectively using the *K*-means clustering algorithm. The total number of psu ($M_{\alpha_0}$) for the list frame, WiFS and LISS III are $M_{1_0} = 6749$, $M_{2_0} = 262000$, $M_{3_0} = 1428000$, respectively. We identify the belongingness of each CCE data points on each SRSD derived spectral clusters. We found that the entire CCE data set is distributed into 6 clusters of WiFS and 4 clusters of LISS III data. We have compared the mean square error (MSE) between psu's (8$^{th}$ row of the table 2) and found that the clusters in the LISS III has the minimum between-psu-MSE (399900 (kg/ha)$^2$) as compared to the WiFS (620700 ((kg/ha)$^2$) and the list frame (871700 (kg/ha)$^2$). We have applied the conventional single-frame-two-stage sampling estimator on each frame separately. We found that the standard error (*S.E.*) for the LISS III frame is minimum (498 kg/ha) as compared to the WiFS (510 kg/ha) and the list frame (626 kg/ha).The estimators of mean and variances used in this table are taken from Singh and Choudhary (1997). These are usual estimators of two-stage sampling.

Table 3 shows the average yield of the state of Haryana for 1997-98 obtained using multiple-frame estimation method for different combinations. We have calculated the *S.E.* of these estimates as the square root of their variances. We found that in the present situation, the multiple-frame combination gives the best estimate of the average yield in respect to their relative efficiency ($R.E. = \frac{\min (S.E.(\bar{y}_{..}))}{S.E.(\bar{y}_{..})}$).

We found that the *R.E.* of the multiple-frame esimator is maximum (equal to 1). We have also measured the *R.E.* of other combinations with respect to this. It is clear from this table that dual frame approach performs better compared to the single frame estimation. The multiple-frame approach using three frames (i.e. "list-WiFS-LISS III" ) perform superior to the single and dual frame estimates.

A comparison of the percentage deviations (PDs) of the different estimators with the Haryana Government′s Estimate of Wheat Yield (HGEWY, 2008) are given in table 3 (column 6). The results show that the single frame estimates with WiFS and LISS III and double frames estimate with WiFS-LISSIII, underestimates the yield as compared to the HGEWY. Whereas single frame with list, double frames with list-WiFS, list-LISSIII and multiple-frame, over-estimates as compared to the HGEWY. Also the multiple-frame estimates has the lowest PD.

We require both the average yield and the area under cultivation for crop production estimation. We derived the acreage under wheat crop from the same satellite data set. Table 4 shows the area under the state of Haryana and the districts Rohtak and Jhajjar as obtained from the satellite data. For estimating the crop area, hyperspectral signatures of wheat are used. The detail methodology can be found in Das and Singh (2009).

## 8. Conclusions

Our approach is different from the *screening* muliple frame agricultural surveys described by Gonzalez-Villalobos and Wallace (1996), where sampling frames are prescreened to remove overlap to form disjoint sets. Here, the sampling elements need not to be the only member of a single frame. The estimator is also different from the estimator proposed by Saxena *et al*. (1984) for multiple-frame samples drawn in multi-stage sampling. Their estimator is an extension of the Hartley's (1962) work for multi-stage sampling. The deficiency of different set of weights for response variable still remains with these estimators. Our estimator is based on the single frame approach of expressing the multiple-frame sampling.

The main contribution in the present work is formulating a general expression of weight given to each sample observation. These weights are function of number of samples drawn and the total number of psu′s and ssu′s. We have calculated these weights based on the belongingness of the sample observations on the list frame and the satellite frames. We have shown that the multiple-frame approach is an efficient technique for estimating the average yield of crop, based on the satellite-remotely sensed data. The approach of estimating average yield using conventional crop cutting experiment data along with satellite-remotely sensed data seems to be most precise in terms of the associated standard errors of the estimates. We have shown that the multiple-frame estimates are better compared to the single and dual frame estimates in terms of their relative efficiencies and percentage deviations.

We have not used the pseudo-maximum likelihood estimator of Lohr and Rao (2006) due to small sample size (in our case it is only 71 CCE data). Theoretically, the pseudo-maximum likelihood estimator outperforms all other esitmators when sample size is large. In the present study we have only explored the single frame estimator without any raking ratio adjustment. In future we would like to experiment with raking ratio, regression and psudo-maximum likelihood estimators.

## Acknowledgments

The authors are grateful to the referees for their critical examination of the paper and very valuable suggessions, which led to considerable improvements to the paper.

## References

ASHTECH, 1993, Z-12 GPS receiver Operating Manual (Santa Clara, CA: Ashtech Ltd).

BANKIER, M. D., 1986, Estimators based on several stratified samples with applications to multiple frame surveys. *Journal of the Ameriacn Statistical Association,* **81**, pp. 1074-1079.

COCHRAN, W. G., (3rd Edition), 1977, *Sampling Techniques,* pp. 292-324, (New York: John Wiley & Sons).

DADHWAL, V. K., PARIHAR, J. S., MEDHAVY, T. A., RAHUL, D. S., and JAISWAL, S. D., 1987, Wheat acreage estimation of Haryana for 1986-87 using Landsat MSS data. Scientific note IRS-UP/SAC/CPR/SN/15/87.

DAS, S. K., and SINGH, R., 2009, Performance of kriging based soft classification using WiFS/ IRS 1D and ground hyperspectral signatures. *IEEE Geosciences and Remote Sensing Letters*, **6(3)**, July 2009, pp. 453-457.

DUDA, R. D., HART, P. E., and STORK, D. G., (Second Edition), 2001, *Pattern Classification and Scene Analysis.* pp. 615-616, (New York: John Wiley & Sons,).

FUKUNAGA, K., (Second Edition), 1990, *Introduction to statistical pattern recognition*, pp. 508-562 (Morgan Kaufmann).

FULLER, W. A., and BURMEISTER, L. F., 1972, Estimators for samples selected from two overlapping frames. in ASA *Proceedings of the Social Statistics Section,* pp. 245-249.

GONZALEZ-VILLALOBOS, A., and WALLACE, M. A., 1996, Multiple frame agricultural surveys. Vols. 1 & 2, Rome: Food and Agriculture Organization of the United Nations.

HARTLAY, H. O., 1962, Multiple frame surveys. *Proceedings of the Social Statistics Section, American Statistical Association*, pp. 203-206.

HARTLAY, H. O., 1974, Multiple frame methodology and selected applications. *Sankhya*, Ser. C, **36**, pp. 99-118.

HGEWY(Haryana Government′s Estimates of Wheat Yield), 2008, National informatics center, *Average yield of principal crops*, Available online at: http://haryana.gov.in/Agriculture/agriculture1.asp.

IDRISI 32, 2001, release 2, *Guide to GIS and Image Processing*, vol. 1, Clark labs, USA, pp. 1-171.

KALTON, G., and ANDERSON, D. W., 1986, Sampling rare populations, *Journal of Official Statistics,* **9**, pp. 747-764.

LOHR, S. L. and RAO, J. N. K., 1997, Jackknife variance estimation in dual frame surveys. Technical Report, Department of Mathematics and Statistics, Carleton University.

LOHR, S. L., and RAO, J. N. K., 2000, Inference in dual frame surveys, *Journal of the American Statistical Association,* **95**, pp. 271-280.

LOHR, S. L., and RAO, J. N. K., 2006, Estimation in multiple-frame surveys, *Journal of the American Statistical Association,* **101**, pp. 1019-1030.

LOHR, S. L., and RAO, J. N. K., 2007, Bootstrap variance estimation in multiple frame surveys. Technical report.

MAGELLAN SYSTEMS CORPORATION, 1999, Magellan GPS 315/320 User Manual (San Dimas, CA: Magellan Systems Corporation).

MATLAB® "The Language of Technical Computing," *Version 7*.0. 4. 365 (R14), 2006.

MUKHOPADHYAY, P., 1998, *Theory and methods of survey sampling,* pp. 237-284, ( New Delhi: Prentice-Hall of India).

NATIONAL AGRICULTURAL STATISTICS SERVICE, 1999, U.S. Equine Inventory Up 1.3 Percent, news release, March 2, available online at h*ttp://www.usda.gov/nass/pubs/pubs.htm*.

PARIHAR, J.S., PANIGRAH, S., DADHWAL, V.K., BHATT, H.P., DASS, N.K., GHOSH, B.K., and BEHRA, D., 1987, Rice acreage estimation in south Orissa using stratified random sampling approach. Scientific note, IRS-UP/SAC/CPR/SN/09/87, 24.

RAO, J. N. K., 1983, H. O. Hartley's contributions to sample survey theory and methods. *The American Statistician,* **37**, pp. 344-350.

SAXENA, B. C., NARAIN, P., and SRIVASTAVA, A. K., 1984, Multiple frame surveys in two stage sampling, *Sankhyā,* Ser. B, **1**, pp. 75-82.

SINGH, D., and CHOUDHARY, F. S.,1997, *Theory and analysis of sample survey designs*, pp. 222-260, ( New Delhi: New age international publishers).

SINGH, R., GOYAL, R., C., SAHA, S. K., and CHHIKARA, R. S., 1992, Use of satellite spectral data in crop yield estimation surveys. *International Journal of Remote Sensing,* **13(14)**, pp. 2583-2592.

SINGH, R., SEMWAL, D. P., RAI, A., and CHHIKARA, R. S., 2002, Small area estimation of crop yield using remote sensing satellite data. *International Journal of Remote Sensing,* **23(1)**, pp. 49-56.

SINGH, R, 2003, Use of satellite data and farmer′s eye estimate for crop yield modeling. *Journal of Indian Society of Remote Sensing,* **56(2)**, pp. 166-176.

SKINNER, C. J., 1991, On the efficiency of raking ratio estimation for multiple frame surveys. *Journal of the American Statistical Association,* **86**, pp. 779-784.

SKINNER, C. J., HOLMES, D. J., and HOLT, D., 1994, Multiple frame sampling for multivariate stratification, *International Statistical Review,* **62**, pp. 333-347

SKINNER, C. J., and RAO, J. N. K., 1996, Estimation in dual frame surveys with complex designs. *Journal of the American Statistical Association,* **91**, pp. 349-356.

SUKHATME, P.V., SUKHATME, B.V., SUKHATME, S., ASHOK, C., (Third Edition), 1984, *Sampling theory of surveys with applications,* pp. 312-369, (Indian society of agricultural statistics, New Delhi).

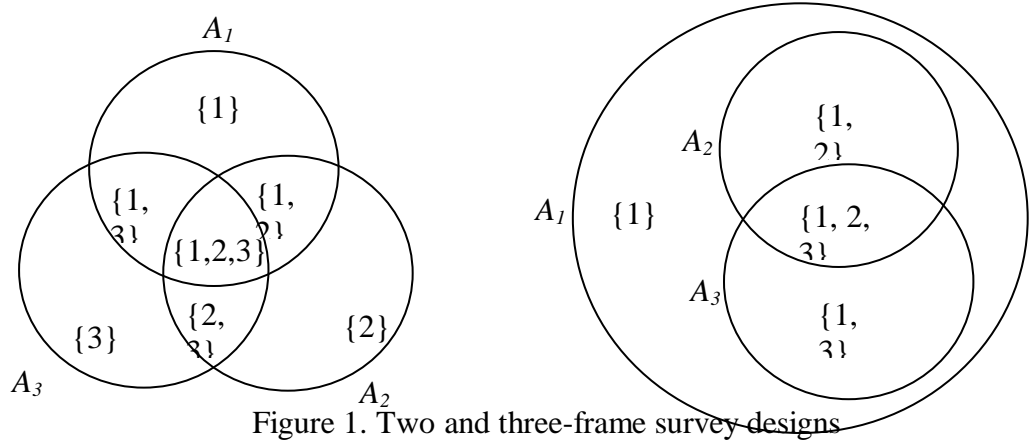

Figure 1. Two and three-frame survey designs

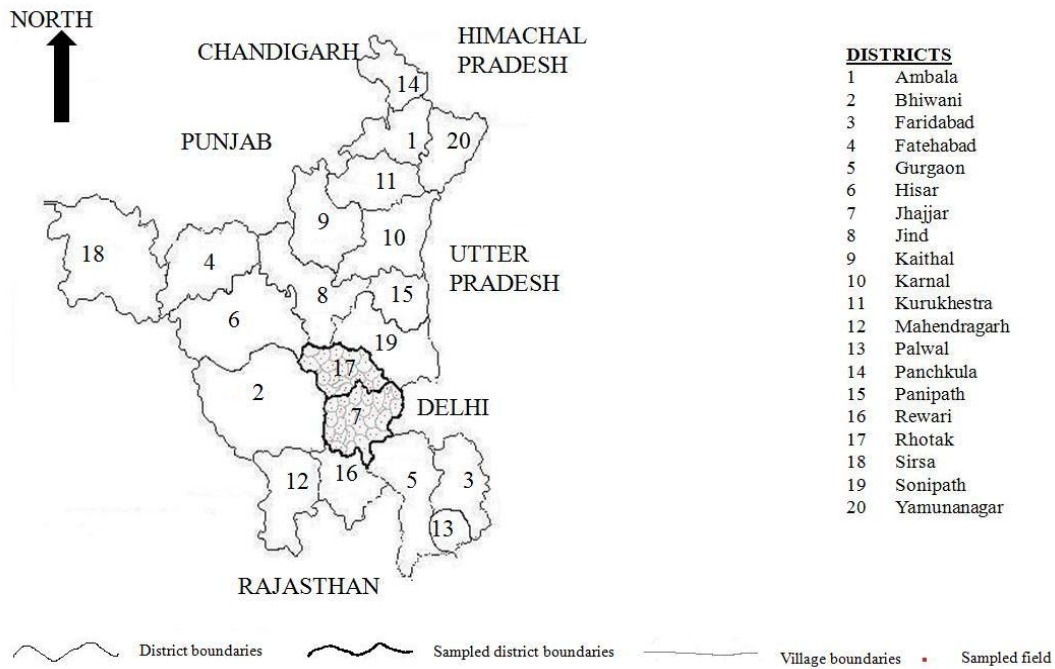

Figure 2. Relationships of different stages of sampling units in a multi-stage sampling for the state of Haryana, India. Two districts (as primary stage units) namely Rhotak and Jhajjar are selected with simple random sampling without replacement (srswor). Random samples of villages (as second stage units) are selected with srswor from each selected district. Two random plots of size 10×5 m$^2$ are taken to measure the yield of each selected village.

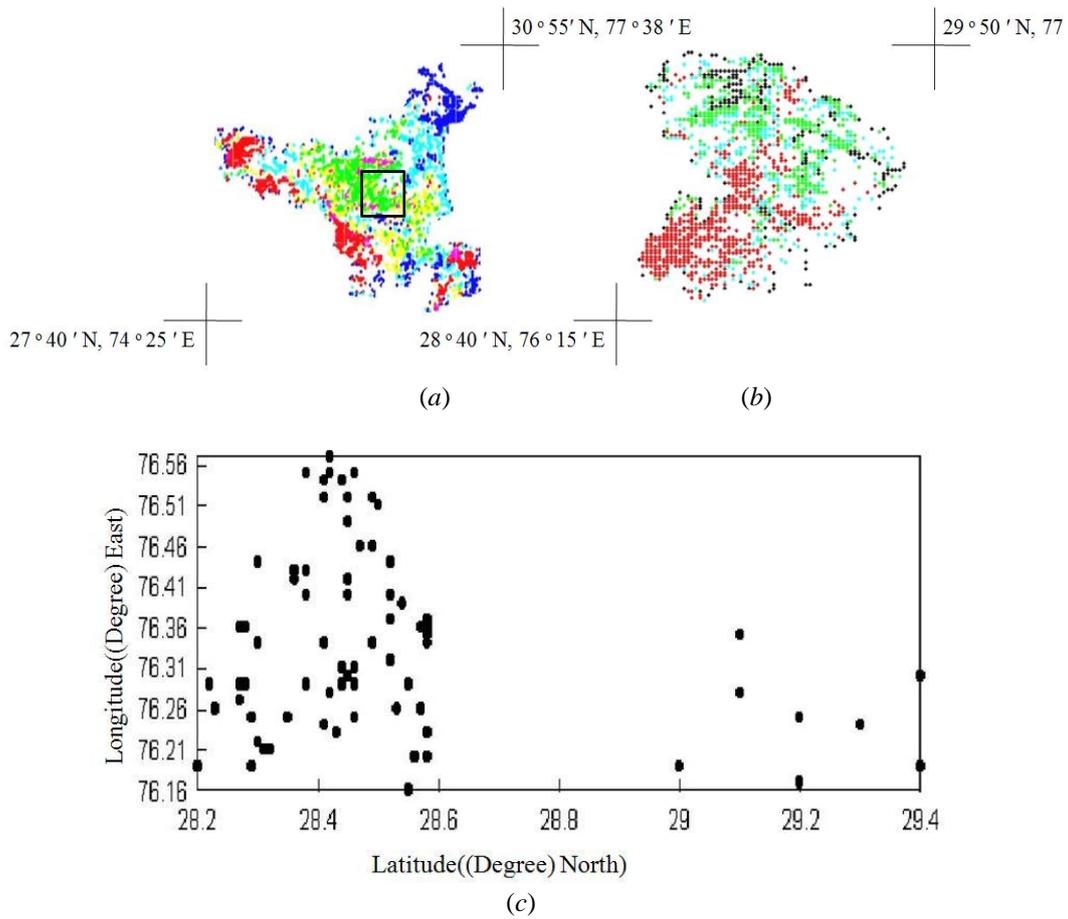

**Figure 4. Spectral clusters of the study area and the crop cutting experiment (CCE) data. The CCE data is assumed to be part of two-stage random samples from the satellite image frames where clusters form the primary stage units and pixels form the second stage units. (*a*) six clusters of WiFS data covering the state of Haryana, India. The rectangular box represents the sampling zone and (*b*) the four clusters from the LISS III data for the districts Rhotak and Jhajjar. (*c*) scatter diagram of the CCE data, the horizontal and vertical axes represent latitude and longitude measuring in degree and minute, the black circles represent the sampled fields.**

Table 4: Crop acreage under wheat as obtained from satellite-remotely sensed data for the districts Rhotak, Jhajjar and the state of Haryana for the Rabi Season (Oct. to Jan.): 1997-98(00 ha).

| Serial No. | Study Area | Estimate of area under wheat using satellite remote sensing data | Area of wheat as reported by The State Govt. |
|---|---|---|---|
| 1. | Rhotak & Jhajjar | 609.25 | 553.00 |
| 2. | Haryana | 4989.17 | 4029.61 |

Table 2: Estimation of the average yield of wheat for the state of Haryana using the list frame and the satellite-remotely sensed data from IRS-1D acquired on 16 $^{th}$ Feb. 1998.

| | Notations | LIST FRAME | WiFS | LISS III |
|---|---|---|---|---|
| Total no. of psu's | $N_\alpha$ | 20 Districts | 500 Clusters | 3000 Clusters |
| Sampled no. of psu's | $n_\alpha$ | 2 Districts | 6 Clusters | 4 Clusters |
| Average no. of ssu's / psu | $\bar{M}_\alpha$ | 337 Villages | 524 | 476 |
| No. of ssu's in *i* th psu | $M_{\alpha_i}$ | $M_{1_1} = 147$, $M_{1_2} = 247$ | $M_{2_1} = 403$, $M_{2_2} = 402$ $M_{2_3} = 325$, $M_{2_4} = 335$ $M_{2_5} = 335$, $M_{2_6} = 397$ | $M_{3_1} = 378$, $M_{3_2} = 286$ $M_{3_3} = 287$, $M_{3_4} = 257$ |
| Sample statistics of ssu's | $m_{\alpha_i}$ | $m_{1_1} = 36$, $m_{1_2} = 35$ | $m_{2_1} = 2$, $m_{2_2} = 3$, $m_{2_3} = 1$ $m_{2_4} = 1$, $m_{2_5} = 1$, $m_{2_6} = 63$ | $m_{3_1} = 1$, $m_{3_2} = 6$, $m_{3_3} = 1$ $m_{2_4} = 63$ |
| Average of *i* th psu (kg/ha) | $\bar{y}_{\alpha_i.}$ | $\bar{y}_{1_1.} = 3440$ $\bar{y}_{1_2.} = 3835$ | $\bar{y}_{2_1.} = 3190$, $\bar{y}_{2_2.} = 3313$ $\bar{y}_{2_3} = 4100$, $\bar{y}_{2_4} = 3260$ $\bar{y}_{2_5} = 2850$, $\bar{y}_{2_6} = 3675$ | $\bar{y}_{3_1.} = 3160$, $\bar{y}_{3_2.} = 3405$ $\bar{y}_{3_3} = 2850$, $\bar{y}_{3_4} = 3675$ |
| Within ssu's mean square error (kg/ha)$^2$ | $s^2_{w_{1i}}$ | $s^2_{w_{1_1}} = 417100$ $s^2_{w_{1_2}} = 460100$ | $s^2_{w_{2_1}} = 101300$, $s^2_{w_{2_2}} = 1447000$ $s^2_{w_{2_3}} = 0$, $s^2_{w_{2_4}} = 0$, $s^2_{w_{2_5}} = 0$, $s^2_{w_{2_6}} = 472400$ | $s^2_{w_{3_1}} = 0$, $s^2_{w_{3_1}} = 718400$ $s^2_{w_{3_1}} = 0$, $s^2_{w_{4_1}} = 472400$ |
| Between psu's mean square error (kg/ha)$^2$ | $s^2_{\alpha_b}$ | $s^2_{1_b} = 871700$ | $s^2_{2_b} = 620700$ | $s^2_{1_b} = 399900$ |
| Average Yield (kg/ha) | $\bar{\bar{y}}_\alpha$ | $\bar{\bar{y}}_1 = 2151$ | $\bar{\bar{y}}_2 = 2372$ | $\bar{\bar{y}}_3 = 2084$ |
| Standard Error (kg/ha) | $S.E.(\bar{\bar{y}}_\alpha)$ | 626 | 510 | 498 |

**Table 3. Comparison of average yield for the state of Haryana for 1997-98 obtained using the multiple-frame approach along with their standard error (S.E.), relative efficiency (R.E.) and percentage deviation.**

| Criteria | Frame Combinations | Average yield ($\bar{\bar{y}}_{..}$) (kg/ha) | S.E. ($\bar{\bar{y}}_{..}$) (kg/ha) | R.E. | Percentage Deviation[1] |
|---|---|---|---|---|---|
| Single Frame | List Frame | 3688.32 | 19.444 | 0.99783 | 0.77377 |
| Single Frame | WiFS | 3394.59 | 116.779 | 0.16614 | -7.25164 |
| Single Frame | LISS III | 3297.91 | 134.072 | 0.14471 | -9.89317 |
| Dual Frame | List Frame and WiFS | 3688.13 | 19.403 | 0.99993 | 0.76857 |
| Dual Frame | List Frame and LISS III | 3688.32 | 19.443 | 0.99790 | 0.77377 |
| Dual Frame | WiFS and LISS III | 3388.79 | 115.765 | 0.16760 | -7.41011 |
| Multiple-Frame | List, WiFS, LISS III | 3688.12 | 19.402 | 1.00000 | 0.76830 |

[1]Calculated as: %Deviation=[HGEWY÷ $\bar{\bar{y}}_{..}$ )×100]-100; where HGEWY=3660 kg/ha (HGEWY,2008)